\documentstyle[12pt,axodraw]{article}
\newcommand{\be}{\begin{equation}}
\newcommand{\ee}{\end{equation}}
\newcommand{\bea}{\begin{eqnarray}}
\newcommand{\eea}{\end{eqnarray}}
\newcommand{\al}{\alpha}

\newcommand{\gm}{\gamma}
\newcommand{\Gm}{\Gamma}

\newcommand{\eps}{\epsilon}
\newcommand{\ep}{\epsilon}

\newcommand{\ka}{\kappa}

\newcommand{\lm}{\lambda}

\newcommand{\dd}{\mbox{d}}

\newcommand{\uq}{\underline{q}}

\newcommand{\um}{\underline{m}}

\newcommand{\nn}{\nonumber}

\begin{document}
\parindent=1.5pc

\begin{titlepage}
\rightline{hep-ph/9703357}
\rightline{March 1997}
\bigskip
\begin{center}
{{\bf
Asymptotic Expansions of Two-Loop Feynman Diagrams \\
in the Sudakov Limit}\\
\vglue 5pt
\vglue 1.0cm
{ {\large V.A. Smirnov}\footnote{E-mail: smirnov@theory.npi.msu.su } }\\
\baselineskip=14pt
\vspace{2mm}
{\it Nuclear Physics Institute of Moscow State University}\\
{\it Moscow 119899, Russia}
\vglue 0.8cm
{Abstract}}
\end{center}
\vglue 0.3cm
{\rightskip=3pc
 \leftskip=3pc
\noindent
Recently presented explicit formulae for asymptotic expansions of 
Feynman diagrams in the Sudakov limit \cite{Sm3} are applied to 
typical two-loop diagrams. For a diagram with one non-zero mass 
these formulae provide an algorithm for analytical
calculation of all powers and logarithms, i.e. coefficients
in the corresponding expansion 
 $(Q^2)^{-2} \sum_{n,j=0} c_{nj} t^{-n} \ln^j t$, with $t=Q^2/m^2$
and $j \leq 4$. Results for the coefficients at several first powers
are presented. For a diagram with two non-zero masses, results
for all the logarithms and the leading power, i.e.
the coefficients $c_{nj}$ for $n=0$ and $j=4,3,2,1,0$ are obtained.
A typical feature of these explicit formulae 
(written through a sum over a specific family of subgraphs
of a given graph, similar to asymptotic expansions for off-shell
limits of momenta and masses) is an interplay between ultraviolet, 
collinear and infrared divergences which represent themselves
as poles in the parameter $\eps = (4-d)/2$ of dimensional 
regularization. In particular, in the case of the second diagram,
that is free from the divergences, individual terms of the asymptotic
expansion involve all the three kinds of divergences resulting in poles, 
up to $1/\eps^4$, which are successfully canceled in the sum.
\vglue 0.8cm}
\end{titlepage}

\section{Introduction}

The simplest explicit formulae for asymptotic expansions of off-shell 
 Feynman diagrams in various limits of momenta and 
masses\footnote{In the large mass limit, one can 
also apply the same `off-shell'
formulae for any values of the external momenta.}
\cite{Go,Ch1,Sm1} (see \cite{Sm2} for a brief review)
have been recently generalized \cite{Sm3,ACVS}
for two on-shell limits:
the limit of the large momenta on the mass shell with the large mass and
the Sudakov limit, with the large momenta on the massless mass shell.

To derive these formulae a method 
based on the construction of a remainder of the expansion and using then
diagrammatic Zimmermann identities \cite{Zi} and 
applied in refs.~\cite{AZ} for operator product expansions within momentum
subtractions and later for diagrammatic and operator expansions
within dimensional regularization and renormalization \cite{Sm1}
has been used.
These general prescriptions were illustrated in 
ref.~\cite{Sm3} through one-loop examples 
(the triangle diagram of Fig.~1a in the case of the Sudakov limit)
and, in ref.~\cite{ACVS},
illustrated and applied for calculation of a typical two-loop diagram
in the case of the first of two limits.

The purpose of this paper is to present similar results in the latter case,
i.e. for the Sudakov limit.
We shall consider two typical vertex two-loop diagrams. The former
is the Feynman integral $F_{1}(p_1,p_2,m)$
corresponding to Fig.~2a with $m_1=\ldots =m_5=0, \;m_6=m$, and
the latter is $F_{2}(p_1,p_2,m)$ with $m_1=\ldots =m_4=0, \;m_5=m_6=m$.
The limit under consideration is $q^2 =  (p_1-p_2)^2 \equiv -Q^2 
\to -\infty$, with $p_1^2 =p_2^2 =0$.
In the case of $F_{1}$, the general prescription will be
used to obtain an algorithm for analytical
calculation of arbitrary coefficients $c^{(1)}_{nj}$
in the corresponding expansion
\be
 F_i(p_1,p_2,m)
\; \stackrel{\mbox{\footnotesize$q^2 \to -\infty$}}{\mbox{\Large$\sim$}} \;
(Q^2)^{-2} \sum_{n=0}^{\infty} \sum_{j=0}^{4}
c^{(i)}_{nj} (m^2/Q^2)^n \ln^j (Q^2/m^2).
\label{exp}
\ee
Results for the coefficients at several first powers
will be presented. For $F_2$, the coefficients
for all the logarithms and the leading power, i.e.
$c^{(2)}_{nj}$ for $n=0$ and $j=4,3,2,1,0$ will be
analytically evaluated.

As in the case of the off-shell explicit formulae \cite{Go,Ch1,Sm1}
the prescriptions of ref.~\cite{Sm3} are written through
a sum over a specific family of subgraphs of a given graph.
A typical feature of the off-shell formulae is an interplay
between ultraviolet and infrared divergences which appear
in individual terms of that sum but are mutually canceled,
provided the initial Feynman integral is finite. It turns out that,
for the Sudakov limit, one meets a more general interplay
between ultraviolet, collinear and infrared divergences which,
at first sight, seem to be of very different nature.
In particular, in the case of $F_2$,
which does not have divergences from the very beginning, 
individual terms of the corresponding asymptotic
expansion involve all three kinds of divergences resulting in poles up to
$1/\eps^4$, with $\eps = (4-d)/2$ parameter of dimensional
regularization \cite{dimreg}. 
However these poles are successfully canceled in the sum
and one obtains expansion (\ref{exp}) in the limit 
$\eps \to 0$.

In the next section we shall briefly recall the general prescriptions
and one-loop example of ref.~\cite{Sm3}. In the following sections
these prescriptions will be applied to the
Feynman integrals $F_{1}$ and $F_{2}$.

\section{General prescriptions and one-loop example}

The asymptotic expansion of a  Feynman integral
$F_{\Gm}(p_1,p_2,m)$, corresponding to a vertex graph $\Gm$,
in the Sudakov limit takes the following explicit form \cite{Sm3}: 
\be
 F_{\Gm}(p_1,p_2,m,\ep)
\; \stackrel{\mbox{\footnotesize$q^2 \to -\infty$}}{\mbox{\Large$\sim$}} \;
\sum_{\gamma}  {\cal M}_{\gm}
 F_{\Gm}(p_1,p_2,m,\ep) \, .
\label{eae1}
\ee        
Here the sum runs over subgraphs $\gm$ of $\Gm$ 
for which at least one of the following conditions holds:

({\em i}) In $\gm$ there is a path between the end-points 1 and 3.
(Three end-points of the diagram are numerated according to
the following order: $p_1, p_2, q=p_1-p_2$.)
The graph $\hat{\gm}$ obtained from $\gamma$ by identifying
the vertices 1 and 3 is 1PI.

({\em ii}) Similar condition with $1 \leftrightarrow 2$.

The pre-subtraction operator ${\cal M}_{\gm}$ is 
defined as a product $\prod_j {\cal M}_{\gm_j}$ of
operators of Taylor expansion acting on 1PI components and cut lines
of the subgraph $\gm$.
Suppose that the above condition ({\em i}) holds and ({\em ii})
does not hold.
Let $\gm_j$ be a 1PI component of $\gm$ and let $p_1+k$ be one
of its external momenta,
where $k$ is a linear combination of the loop momenta.
(We imply that the loop momenta are chosen
in such a way that $p_1$ flows through all $\gm_j$ and the
corresponding cut lines).
Let now $\uq_j$ be other independent external momenta of $\gm_j$.
Then the operator $\cal M$ for this component is defined as
${\cal T}_{k-((p_1 k)/(p_1p_2)) p_2,\uq_j, \um_j } \,$,
where $\um_j$ are the masses of $\gm_j$.
In other words, it is the operator of Taylor expansion in $\uq_j$ and
$\um_j$
at the origin and in $k$ at the point
$\tilde{k} =\frac{(p_1 k)}{(p_1p_2)} p_2$ (which depends on $k$ itself).

 For the cut lines the same prescription is adopted.
If $p_1 +k$ is the momentum of the cut line, then the
corresponding operator acts as
$\left. {\cal T}_{\ka} \frac{1}{\ka (k_l^2-m_l^2) +2 p_1 k }
\right|_{\ka=1} \, .$
If both conditions ({\em i}) and ({\em ii}) hold the corresponding
operator performs Taylor expansion in the mass and the external momenta
of subgraphs (apart from $p_1$ and $p_2$).

 For example, in the case of the one-loop triangle diagram of Fig.~1a,
\be
 F_{\Gm}(p_1,p_2,m,\ep)
= \int \frac{\dd^dk}{(k^2-2 p_1 k) (k^2-2 p_2 k) (k^2-m^2)} \, ,
\label{triangle}
\ee
this family of subgraphs is shown in Fig.~1.
Besides the graph itself (a), these are two subgraphs consisting,
respectively, of the left and the right line (b and c).
The subtraction operator for (a) expands the propagator with the mass
$m$ in Taylor series in $m$, and the subtraction operators ${\cal M}_{\gm}$
for (b) and (c) act at the first and, respectively, the second
factor in the integrand in (\ref{triangle}) as follows:
\be
{\cal M}^a_i \frac{1}{k^2 - 2 p_i k}
= \sum_{j=0}^a \frac{(k^2)^j}{(- 2 p_i k)^{j+1}} \, , \;\; i=1,2 \, .
\label{Ma12}
\ee

All the resulting one-loop integrals are easily evaluated for any order
of the expansion.
The operator ${\cal M}_0$ gives the following contribution
at $\eps\neq 0$:
\be
{\cal M}_0 F_{\Gm} = -i \pi^{d/2} \frac{1}{(-q^2)^{1+\eps}}
\sum_{n=0}^{\infty}
\frac{\Gm(n+1+\eps)\Gm(-n-\eps)^2}{\Gm(1-n-2\eps)}
\left(\frac{m^2}{q^2}\right)^n \, .
\label{M00}
\ee
The terms ${\cal M}_1F_{\Gm}$ and ${\cal M}_2F_{\Gm}$ are not 
individually regulated by dimensional regularization but their 
sum exists for general $\eps\neq 0$:
\bea
\left( {\cal M}_1 + {\cal M}_2 \right) F_{\Gm} = i \pi^{d/2}
\frac{1}{q^2 (m^2)^{\eps}} \Gm(\eps) \Gm(1-\eps)
\sum_{n=0}^{\infty} \frac{1}{n! \Gm(1-n-\eps)}
\nn \\ \times
\left[ \ln (Q^2/m^2) + \psi(\eps) + \psi(n+1) - \psi(1-\eps)
- \psi (1-n-\eps) \right]
\left(\frac{m^2}{q^2}\right)^n \, .
\label{M012}
\eea
Here $\psi$ is the logarithmic derivative of the gamma function.
In the limit $\eps\to 0$, one arrives at the known result:
\be
F_{\Gm} \left. \right|_{\eps=0}
\; \stackrel{\mbox{\footnotesize$q^2 \to -\infty$}}{\mbox{\Large$\sim$}}
= - \frac{i\pi^2}{q^2} \left[
\mbox{Li}_2\left(\frac{1}{t}\right) + \frac{1}{2} \ln^2 t
- \ln t \ln(t-1) -\frac{1}{3} \pi^2 \right] \, ,
\label{tri-res}
\ee
where Li$_2$ is the dilogarithm and $t=Q^2/m^2$.

\section{Diagram with one non-zero mass}

The two-loop Feynman integrals under consideration can be written as
\bea
 F_i(p_1,p_2,m,\ep) \nn \\
= \int  \int \frac{\dd^dk \dd^dl}{(l^2-2 p_1 l) (l^2-2 p_2 l)
(k^2-2 p_1 k) (k^2-2 p_2 k) (k^2-m_{5i}^2) ((k-l)^2-m^2)} \, ,
\label{FINT}
\eea
where $m_{51}=0$  and $m_{52}=m$.
Let us apply the above prescriptions to them.
The family of subgraphs that give non-zero contributions
is shown in Fig.~2(a--f). Let us begin with the first case, when
subgraphs (c) and (d) also happen to give zero contributions.

The contribution of Fig.~2a is obtained by expanding the 
massive propagator in the geometric series in $m$. 
In the leading order, this is  Fig.~2a with pure zero masses which
was calculated in \cite{Gons,vanN,KL}:
\be
\frac{(i\pi^{d/2})^2}{(Q^2)^{2+2\eps}} \frac{1}{\eps}
\left\{ \frac{1}{2 \eps} G_2(2,2) G_3(2+\eps,1,1)
- G_2(2,1) \left[ \frac{1}{\eps} G_3(2,1,1+\eps) + G_3(1,1,1)\right] 
\right\} \, ,
\label{leading(a)}
\ee
where  $G_2(a,b)$ and $G_3(a,b,c)$ are expressed
through gamma functions and give, respectively,
the value of the general one-loop propagator-type integral and
triangle integral with the given kinematics, in the case
of arbitrary indices of lines.

All the resulting Feynman integrals in arbitrary order of the
expansion are evaluated by integration by parts \cite{IBP}
(as in \cite{KL})
and are expressed, for general $\eps$, in terms of gamma functions.

To write down the contribution of Fig.~2b 
it is necessary to expand the initial diagram, with the 6th line omitted,
in Taylor series (under the sigh of the integral over the loop
momenta) in the loop momentum flowing through this line
in the whole diagram and then insert the result 
into the reduced diagram which is 
the tadpole generated by the 6th line. This subgraph Fig.~2b 
contributes to the expansion starting from the next-to-leading order, 
$m^2/(Q^2)^3$.
The calculation of this contribution is rather simple and reduces
to successive application of the one-loop integration formulae,
with the $G$-functions $G_2$ and $G_3$ in the result. 

What is more non-trivial is the evaluation of the contributions
of the subgraphs of Fig.~2e and f. 
The contribution of Fig.~2e is obtained by expansion
of the propagators $1/(l^2-2 p_1 l)$ and $1/(k^2-2 p_1 k)$ in
geometrical series in $l^2$ and $k^2$, respectively:
\bea
\sum_{N_1,N_3} 
\int  \int \frac{\dd^dk \dd^dl 
(-1)^{N_1+N_3}(l^2)^{N_1} }{(-2 p_1 l)^{1+N_1} (l^2-2 p_2 l)
(-2 p_1 k)^{1+N_3} (k^2-2 p_2 k) (k^2)^{1-N_3} ((k-l)^2-m^2)}.
\label{subgraph_e}
\eea
However these contributions are not individually
regularized by dimensional regularization so that let us introduce,
temporarily, analytic regularization into the lines number
3 and 4, with $\lm_3 \neq \lm_4$, then calculate the integrals
involved and switch off the regularization in the sum.

Thus the problem reduces to calculation of the following
family of integrals:
\bea
J(N_1,\ldots ,N_7) = 
\int  \int \frac{\dd^dk \dd^dl (l^2)^{N_7}}{(-2 p_1 l)^{1+N_1} 
(l^2-2 p_2 l)^{1+N_2} (-2 p_1 k)^{1+N_3} (k^2-2 p_2 k)^{1+N_4}} 
\nn \\ \times
\frac{1}{(k^2-m_5^2)^{N_5} ((k-l)^2-m^2)^{N_6} } \, , 
\label{Je+f}
\eea
with integer $N_6$ and $N_7$.
Using integration by parts \cite{IBP} it is possible to express any
integral with  $N_7>0$ through a linear combination of integrals
with  $N_7=0$.

 For integral (\ref{Je+f}) with $N_7=0$, let us use the $\al$-parametric
representation. Then, in the resulting 6-fold integral, it is
possible to apply the Mellin-Barnes representation in such a way
that the internal integration over $\al$-parameters can be
performed explicitly and we are left with the following
representation:
\bea
J(N_1,N_2,N_3+\lm_3,N_4+\lm_4,N_5,N_6,0)
= \frac{(i\pi^{d/2})^2}{(Q^2)^{2+N_1+N_3+\lm_3}
(m^2)^{2 \eps+N_2+N_4+N_5+N_6 +\lm_4}}
\nn \\
\times \frac{\Gm(-\eps-N_3-N_5-\lm_3) \Gm(-\eps-N_4-N_5-\lm_4)
\Gm(2 \eps+N_2+N_4+N_5+N_6+\lm_4)}{
\Gm(1+N_2) \Gm(1+N_4+\lm_4) \Gm(1+N_5) \Gm(1-\eps-N_5)
\Gm(-N_1-N_3-\eps -\lm_3)}
\nn \\ \times
\frac{1}{2\pi i}\int_{\cal C} \dd s \Gm(1+N_5+s)
\frac{\Gm(s-\eps-N_2+1)}{\Gm(s-\eps-N_3-\lm_3+1)}
\nn \\ \times
\Gm(s-N_1-N_3-\eps -\lm_3)
\Gm(\eps+N_2+N_4+\lm_4-s) \Gm(-s) \,.
\eea
The contour of integration is chosen in the standard way: the
IR poles are to the left and the UV poles are to the right. 
The singularities at $\lm_3 \to \lm_4$ are due to the fact
that some UV poles become identical with IR poles.
By shifting the contour of integration these pole contributions
are explicitly picked up. The rest of the integral is finite
in the limit  $\lm_3,\lm_4 \to 0$, and
the resulting integrals are easily evaluated by 
recurrence relations and finally using well-known integrals
with four gamma functions.

 Following this procedure, any integral 
that is present in the contribution of Fig.~2e and~f
can be analytically evaluated, for arbitrary $\eps$.
 For example, the leading contribution takes the form
\be
- \frac{(i\pi^{d/2})^2}{(Q^2)^{2} (m^2)^{2 \eps}}
\Gm(-\eps) \Gm(\eps) \Gm(2\eps)
[ L + \psi(2\eps) + \psi(1+\eps)-2 \psi(1)] \, ,
\label{leadinga1eps}
\ee
where $L=\ln Q^2/m^2$.
To avoid the Euler constant in the results we present
several first terms of the expansion of the function
$F_1 \cdot  (Q^2)^{2}/(i\pi^{d/2} \Gm(1+\eps))^2 =
\sum_{n=0}^{\infty} C_n t^{-n}$:
\bea     
C_0  & = &
\left(  {\textstyle{1\over2}}L^2 
+ {\textstyle{1\over3}}\pi^2 \right) 
{\textstyle{1\over\eps^2}}
- \left(  {\textstyle{1\over3}}L^3
+ {\textstyle{1\over6}}\pi^2 L  - 3 \zeta(3) \right) 
{\textstyle{1\over\eps}}
\hspace{3cm}  \nn \\ &&
+ {\textstyle{1\over6}}L^4 + {\textstyle{1\over3}}\pi^2 L^2   
- 11 \zeta(3) L + {\textstyle{61\over360}} \pi^4  \, ,
\label{C0} \nn \\
C_1 & = &
-  \left(L+ 1 \right) 
{\textstyle{1\over\eps^2}}
+ \left( L^2 - L  + {\textstyle{1\over6}}\pi^2 \right) 
{\textstyle{1\over\eps}}
\hspace{3cm} \nn \\ &&
-  {\textstyle{2\over3}} L^3  + 3 L^2 - 6 L -  {\textstyle{2\over3}} \pi^2 L
-26  + {\textstyle{1\over2}} \pi^2  + 11 \zeta(3)  \, ,
\label{C1} \nn \\
C_2 & = &
- \left( {\textstyle{1\over2}} L + {\textstyle{1\over4}} \right)
{\textstyle{1\over\eps^2}}
+ \left({\textstyle{1\over2}} L^2 - {\textstyle{7\over4}} L
- {\textstyle{5\over4}} + {\textstyle{1\over12}} \pi^2 \right)
{\textstyle{1\over\eps}}
\hspace{3cm}  \nn \\&&
- {\textstyle{1\over3}}L^3 + {\textstyle{13\over4}} L^2
- {\textstyle{5\over2}} L - {\textstyle{1\over3}} \pi^2 L  
-{\textstyle{117\over8}}  + {\textstyle{17\over24}} \pi^2
+ {\textstyle{11\over2}} \zeta(3) \, ,
\label{C2} \nn \\
C_3 & = &
- \left( {\textstyle{1\over3}} L + {\textstyle{1\over9}} \right)
{\textstyle{1\over\eps^2}}
+ \left( {\textstyle{1\over3}}L^2 - {\textstyle{29\over18}} L
- {\textstyle{3\over4}} + {\textstyle{1\over18}} \pi^2 \right)
{\textstyle{1\over\eps}}
\hspace{3cm}  \nn \\&&
- {\textstyle{2\over9}} L^3 + {\textstyle{17\over6}} L^2
- {\textstyle{35\over9}} L - {\textstyle{2\over9}} \pi^2 L
- {\textstyle{890\over81}} + {\textstyle{23\over36}} \pi^2
+ {\textstyle{11\over3}} \zeta(3)
\label{C3} \,.
\eea     
These results have been obtained with the help of the
{\sl Mathematica} system \cite{math}.

\section{Diagram with two non-zero masses}

In the case of the second Feynman integral that we
consider there are contributions from all the subgraphs of
 Fig.~2. The contribution of Fig.~2a is obtained by expanding the propagators
of the 5th and the 6th lines in the geometric series in $m$.
As in the previous case the resulting integrals are calculated by recurrence
relations based on integration by parts.
The leading contribution has the same form (\ref{leading(a)}).
The contribution of Fig.~2b in arbitrary order also poses no difficulties, 
with a result expressed through the $G$-functions $G_2$ and $G_3$.

Consider now  Fig.~2c and~d. According to the general prescription,
the contribution of Fig.~2c is written through the product of
the expansion of the propagator $1/(k^2-2 p_1 k)$ in geometrical series
with respect to $k^2$ and an expansion of the lower
triangle subdiagram which is a function 
$f(q_1^2,q_2^2,q_3^2)$ of the following external momenta squared:
$q_1^2= k^2-2 p_1 k, \, q_2^2=k^2-2 p_2 k $ and 
$q_3^2=q^2$. So, this second expansion of $f$ is in Taylor series
(again in the sense of the expansion under the sign of integral)
in $q_2^2$, with the subsequent expansion of the result,
as a function of $k^2-2 p_1 k$ in $k^2$.
All the derivatives of that triangle in $q_2^2$ at  $q_2^2=0$
are calculated by recurrence relations based on integration
by parts (see, e.g., results for the first two derivatives in \cite{DOT}).

This product of the expansion of the 4th line and the lower
triangle is then integrated with the product of other
propagators (number 4 and 6), using simple one-loop integration
formulae. However this contribution of
 Fig.~2c alone is not regularized dimensionally so that
it is natural to consider the sum of both contributions of 
 Fig.~2c and~d. Still to handle these terms individually, one
can introduce a temporal analytic regularization, into the lines
number 3 and 4, calculate them and switch off the analytic
regularization in the sum.

 Following this procedure, the contribution of Fig.~2c 
and~d can be evaluated in terms of gamma functions in every order 
of the expansion, for arbitrary $\eps$.
 For example, the leading, $1/(Q^2)^2$, contribution takes the
form
\be
\frac{(i\pi^{d/2})^2}{(Q^2)^{2+ \eps} (m^2)^{ \eps}}
\frac{\Gm(\eps)^2 \Gm(1-\eps)^2}{\eps\Gm(1-2\eps)}
\left\{ -2 \frac{\Gm(-\eps)^2}{\Gm(-2\eps)}
+  \ln t + \psi(\eps) -2 \psi(-\eps) + \psi(1) \right\} \, .
\label{leading(c+d)}
\ee

To calculate the contribution of  Fig.~2e and~f it is reasonable
to use the Mellin-Barnes representation for the 5th propagator
and reduce the problem to calculation 
of the corresponding contribution for the first of our integrals,
in the case when this line is analytically regularized.
In particular, the leading, $1/(Q^2)^2$, contribution is obtained 
as the following integral of the 
analytically regularized version of eq.~(\ref{leadinga1eps}):
\bea
-\frac{(i\pi^{d/2})^2}{(Q^2)^{2+ \eps} (m^2)^{ \eps}}
\frac{1}{2\pi i}\int_{\cal C} \dd s \Gm(2\eps+s) \Gm(\eps+s)
\Gm(-\eps-s) \Gm(-s)
\nn \\ \times 
[\ln t + \psi(2\eps+s) + \psi(1+\eps+s) -2 \psi(1+s)] \, .
\label{leadinga2eps}
\eea
The contour ${\cal C}$ is chosen in the standard way, with the 
qualification that the pole $s=-\eps$ (which is both UV and IR) 
is to the right of it. The pole and the finite part of 
eq.~(\ref{leadinga2eps}) are evaluated separately
and give the following result written up to $\eps^0$:
\bea
-\frac{(i\pi^{d/2})^2}{(Q^2)^2 (m^2)^{2\eps}}
\left\{ 
2\zeta(3) \ln t + {\textstyle{\pi^4\over30}} 
\right. \nn \\ \left. 
+ \Gm(-\eps) \Gm(\eps) \Gm(2\eps) 
\left[ \ln t + \psi(1-\eps) -2\psi(1-2\eps) - \psi(-\eps)
+ \psi(\eps) + \psi(2\eps) \right] \right\} \, .
\label{leadinga2eps0}     
\eea

Now, the leading order of the asymptotic expansion 
of the second of our Feynman integrals is given explicitly
by the sum of three terms corresponding to Fig.~2a, 
 Fig.~2(c\&d) and Fig.~2(e\&f), respectively, (\ref{leading(a)}),
(\ref{leading(c+d)}) and (\ref{leadinga2eps0}).
Note that the first term involves infrared and collinear divergences,
the third term involves ultraviolet and collinear divergences,
and the second term possesses all the three kinds of the divergences.
However  all the poles, up to $1/\eps^4$, 
are canceled in the sum and we get
\be
 F_2 (Q^2,m^2)
\; \stackrel{\mbox{\footnotesize$Q^2 \to \infty$}}{\mbox{\Large$\sim$}} \;
\frac{-\pi^4}{(Q^2)^2} \left(
\frac{1}{24} \ln^4 t + \frac{\pi^2}{3} \ln^2 t
- 6 \zeta(3) \ln t + \frac{31 \pi^4}{180}
\right) \,.
\label{leading2}
\ee
These results are in a good agreement with numerical MC calculations:
e.g., there is 2\% accuracy at $t=50$. 
However this accuracy is achieved if all the terms in the
parenthesis are taken into account. In other words, it is
necessary to exhaust all the powers of the logarithm until
the remainder is determined by the next power, $m^2 /(Q^2)^3$.

 The calculation of the contribution of  Fig.~2e and~f in arbitrary
order of the expansion happens to be more elaborated and will
be reported in  future publications as well as 
details of the present calculations, similar results 
for non-planar diagrams, with subsequent application of 
the new information obtained in two loops to 
extend well-known results on asymptotic behaviour in the Sudakov
limit in QED and QCD \cite{Sud,Sud1,Co,Kor}.

\vspace{2mm}

This research has been supported by the Russian Foundation for Basic
Research, project 96--01--00654 and by INTAS, project 93--0744.
\vspace{2mm}

{\em Acknowledgments.}
I am grateful to K.G.~Chetyrkin, A.I.~Davydychev, 
K.~Melnikov and J.B.~Tausk for helpful discussions, and 
to J.~Fleischer for numerical results for the Feynman integral $F_2$.

\begin {figure} [htbp]
\begin{picture}(400,90)(-100,10)

\Line(10,100)(60,100)
\Line(60,100)(35,70)
\Line(35,70)(10,100)
\ArrowLine(35,50)(35,70)
\ArrowLine(10,100)(10,120)
\ArrowLine(60,120)(60,100)

\Text(35,45)[]{$p_1-p_2$}
\Text(10,125)[]{$p_1$}  
\Text(60,125)[]{$p_2$} 
\Text(35,25)[]{$(a)$}
\Text( 25, 70)[]{1}
\Text( 50, 70)[]{2}
\Text( 35, 110)[]{3}

\Vertex(10,100){1.5}
\Vertex(60,100){1.5}
\Vertex(35,70){1.5}

\Line(145,70)(120,100)

\Vertex(120,100){1.5}
\Vertex(145,70){1.5}

\Text(145,25)[]{$(b)$}

\Line(250,100)(225,70)

\Vertex(250,100){1.5}
\Vertex(225,70){1.5}

\Text(225,25)[]{$(c)$}

\end{picture}
\caption {(a) One-loop vertex diagram. (a)--(c)
Subgraphs contributing to the asymptotic expansion
of the diagram (a) in the Sudakov limit.}
\label{1l}

\vspace{2cm}
  
\begin{picture}(400,160)(-35,0)

\Line(10,100)(60,100)
\Line(10,150)(60,150)
\Line(10,100)(10,150)
\Line(60,150)(60,100)
\Line(60,100)(35,70)
\Line(35,70)(10,100)
\ArrowLine(35,50)(35,70)
\ArrowLine(10,150)(10,170)
\ArrowLine(60,170)(60,150)


\Text(35,45)[]{$p_1-p_2$}

\Text(10,175)[]{$p_1$}
\Text(60,175)[]{$p_2$}
\Text(35,25)[]{$(a)$}
\Text(  0,118)[]{3}
\Text( 25, 70)[]{1}
\Text( 73,118)[]{4}
\Text( 50, 70)[]{2}
\Text( 35,140)[]{5}
\Text( 35, 90)[]{6}

\Vertex(10,100){1.5}
\Vertex(60,100){1.5}
\Vertex(10,150){1.5}
\Vertex(60,150){1.5}
\Vertex(35,70){1.5}

\Line(120,150)(170,150)
\Line(120,100)(120,150)
\Line(170,150)(170,100)
\Line(145,70)(120,100)
\Line(170,100)(145,70)

\Vertex(120,100){1.5}
\Vertex(170,150){1.5}
\Vertex(145,70){1.5}
\Vertex(120,150){1.5}
\Vertex(170,100){1.5}

\Text(145,25)[]{$(b)$}


\Line(230,100)(230,150)
\Line(255,70)(230,100)
\Line(280,100)(255,70)

\Line(230,100)(280,100)

\Vertex(230,100){1.5}
\Vertex(230,150){1.5}
\Vertex(280,100){1.5}
\Vertex(255,70){1.5}

\Text(255,25)[]{$(c)$}


\Line(360,150)(360,100)
\Line(335,70)(310,100)
\Line(360,100)(335,70)

\Line(310,100)(360,100)

\Vertex(310,100){1.5}
\Vertex(360,100){1.5}
\Vertex(360,150){1.5}
\Vertex(335,70){1.5}

\Text(335,25)[]{$(d)$}

\end{picture}

\begin{picture}(400,140)(-35,12)


\Line(135,100)(135,150)
\Line(160,70)(135,100)
\Text(160,25)[]{$(e)$}
\Vertex(135,100){1.5}
\Vertex(135,150){1.5}
\Vertex(160,70){1.5}


\Line(265,150)(265,100)
\Line(265,100)(240,70)
\Text(240,25)[]{$(f)$}
\Vertex(265,100){1.5}
\Vertex(265,150){1.5}
\Vertex(240,70){1.5}

\vspace{-10pt}

\end{picture}
\caption {(a) A typical two-loop vertex diagram. (a)--(f) 
Subgraphs contributing to the asymptotic expansion
of the diagram (a) in the Sudakov limit.}
\label{2l}
\end{figure}

\end{document}